\documentclass[aps, prl, twocolumn, nofootinbib, superscriptaddress]{revtex4-1}
\usepackage{epsfig}
\usepackage{amsmath}
\usepackage{amssymb}
\usepackage{subfigure}
\usepackage{bm}   
\usepackage{verbatim} 
\usepackage{natbib}

\begin{document}

\title{Frozen Vacuum}
\date{\today}
\author{Raphael Bousso}
\affiliation{Center for Theoretical Physics and Department of Physics\\
\ \  University of California, Berkeley, CA 94720-7300, U.S.A.}
\affiliation{Lawrence Berkeley National Laboratory, Berkeley, CA 94720-8162,
  U.S.A.}
\begin{abstract}
Modes just outside the horizon of a typical old black hole are thermally entangled with distant Hawking radiation. This precludes their entangled purity with interior modes, leading to a firewall. Identifying the interior with the distant radiation (``$A=R_B$'', ``ER $=$ EPR'') can resolve the entanglement conflict. But the map must adjust for any interactions, or else the firewall will reappear if the Hawking radiation scatters off the CMB. With a self-correcting map, an infalling observer is unable to excite the vacuum near the horizon. This allows the horizon to be locally detected and so violates the equivalence principle.
\end{abstract}
\maketitle

Let $b$ be an outgoing Hawking mode in the near-horizon zone (the ``zone'') of a large black hole.  By unitarity of the S-matrix, this mode is either pure or else purified by the remainder of the entire Hawking radiation. Some of this radiation is yet to be emitted, so the exact purification of $b$ may require degrees of freedom associated with the black hole.  This holds for any black hole, young or old. 

Here I will consider an old black hole, whose entropy is smaller than the entropy of the ``early'' radiation that it has already emitted.  I will assume that the quantum state of the black hole-radiation system is typical, in the sense of Ref.~\cite{Pag93,HayPre07}.\footnote{The present argument complements~\cite{Bou13,Bou13a}, which noted that firewalls are necessarily present if the zone has less than thermal entropy; such states are atypical but form a complete basis.} In such a state, $b$ has large entropy, $S_b\sim O(1)$, but can be purified by the early radiation alone. More precisely, there exists a (highly nonunique) scrambled subsystem $e_b$ of the early radiation such that the von Neumann entropy $S_{be_b}$ is exponentially small~\cite{HayPre07}. In the infalling vacuum, $b$ would need to be purified by an interior mode $\tilde b$: $S_{b\tilde b}\ll 1$, but subadditivity excludes this: $S_b\leq S_{b\tilde b}+S_{be_b}$~\cite{AMPS}.

In order to avoid a firewall at the horizon, one could identify the interior partner $\tilde b$ with some $e_b$ or with the exact purification $\hat b$. This reduces the inconsistent double entanglement to a consistent single entanglement. An out-state-dependent mapping is necessary to ensure that $b \tilde b$ will not just be entangled, but in a particular entangled state, the vacuum state. This type of map is called $A=R_B$~\cite{Bou12c,Sus13,HarHay13,Bou13} or ER=EPR~\cite{MalSus13}, or ``donkey map''~\cite{Bou13a}. It is nonstandard~\cite{Bou12c,AMPSS}, and so already faces a number of challenges;\footnote{If $e_b$ is transported into the black hole, then $\tilde b$ and $e_b$ can be simultaneously and independently accessed and so cannot be identified. The extraction of $e_b$ from the early radiation is computationally challenging~\cite{HarHay13}, but even the coherent transport of a computational qubit $e$ into the interior leads to nonvanishing commutators at spacelike separation~\cite{AMPSS}. The difficulty I will describe here does not require coherent transport of $e$ into the interior; a classical record suffices. Moreover, the sterility of the vacuum at the horizon is physically vivid and manifestly conflicts with the equivalence principle.} moreover, no donkey map exists for out-states where $b$ is less than thermally entangled~\cite{Bou13,MarPol13,Bou13a}. Here I will identify a different challenge: the donkey map makes it impossible for an infalling observer to excite $b$, in violation of the equivalence principle. 

\paragraph{Spread purification} I begin by assuming that $e_b$ can be rapidly unscrambled and measured. For simplicity I will ignore the nonuniqueness of $e_b$ and treat the state of $be_b$ as exactly pure. I will later consider the case where the purification of $b$ cannot be computationally accessed, and where there may be ambiguities or redundancies in $e_b$; this will not change the conclusion. 

We can write the quantum state of $be_b$ as
\begin{equation} 
|\psi\rangle_{be_bp} \propto  |0\rangle_p \otimes \sum_{n=0}^\infty x^n |n\rangle_b |n\rangle_{e_b}~.
\label{eq-1}
\end{equation}
I have included a pointer $p$ which has not yet interacted with either $b$ or $e_b$. The pointer could be a measurement apparatus or any other environment. The states of $e_b$ have been labeled so as to make the donkey map look simple: 
\begin{equation}
|0\rangle_{e_b} \to |0\rangle_{\tilde b}~,~~ |1\rangle_{e_b} \to |1\rangle_{\tilde b},~~\ldots
\label{eq-olddonkey}
\end{equation}
I suppress normalization factors (here, $\sqrt{1-x^2}$). For modes with Killing frequency of order the Hawking temperature, $x\equiv e^{-\beta\omega/2}$ is of order one. To be concrete, I will assume that the black hole is billions of light-years in size.

Suppose that the pointer measures $e_b$ in the above basis. Now the state is
\begin{equation} 
|\psi\rangle_{be_bp} \propto \sum_{n=0}^\infty x^n |n\rangle_b |n\rangle_{e_b} |n\rangle_p
~.
\label{eq-2}
\end{equation}
The unitary evolution from (\ref{eq-1}) to (\ref{eq-2}) is called a pre-measure\-ment. 

A realistic measuring apparatus is not well-insulated from the environment. We could include the environment explicitly, but the same effect can be captured by considering the pointer to be the environment. Once the environment is excluded from the description, the system will be in a mixed state; the wave function collapses. This is seen by taking a partial trace over $p$ in the state (\ref{eq-2}), which yields
\begin{equation}
\rho_{be_b}=(1-x^2) \sum_{n=0}^\infty x^{2n} |n\rangle_b\, |n\rangle_{e_b}\,\,\mbox{}_{e_b} \langle n|\, \mbox{}_b \langle n|~.
\label{eq-decohere}
\end{equation}
But this state cannot be mapped to $|0\rangle_{b\tilde b}$ under any map from $e_b$ to $\tilde b$, because it is not pure.

Firewalls are equally troubling whether the Hawking radiation interacts with an environment or not; generically, it will. To preserve the infalling vacuum, the donkey map must keep track of all environmental degrees of freedom that the radiation interacts with. It must adjust to the new state, Eq.\,(\ref{eq-2}). With the updated donkey map,
\begin{equation}
|0\rangle_{e_b} |0\rangle_p \to |0\rangle_{\tilde b}~,~~|1\rangle_{e_b} |1\rangle_p \to |1\rangle_{\tilde b}~~\ldots~,
\label{eq-donkey}
\end{equation}
the state (\ref{eq-2}), too, becomes the infalling vacuum:
\begin{equation} 
|0\rangle_{b\tilde b} \propto \sum_{n=0}^\infty x^n |n\rangle_b |n\rangle_{\tilde b} ~.
\label{eq-3}
\end{equation} 

To summarize, for the donkey map to restore the infalling vacuum, the pre-measure\-ment (\ref{eq-2}) cannot be treated as a decohered measurement, Eq.~(\ref{eq-decohere}). Instead, the environment must be included in the Hilbert space that the map acts on, so that interactions merely spread the purification of $b$ to include new degrees of freedom. Note that this must be done whether or not the infalling observer has any idea that an interaction has taken place, let alone any control over the environment. I will now show that this freezes the vacuum at the horizon.
\begin{figure}[tbp]\centering
\includegraphics[width=.4 \textwidth]{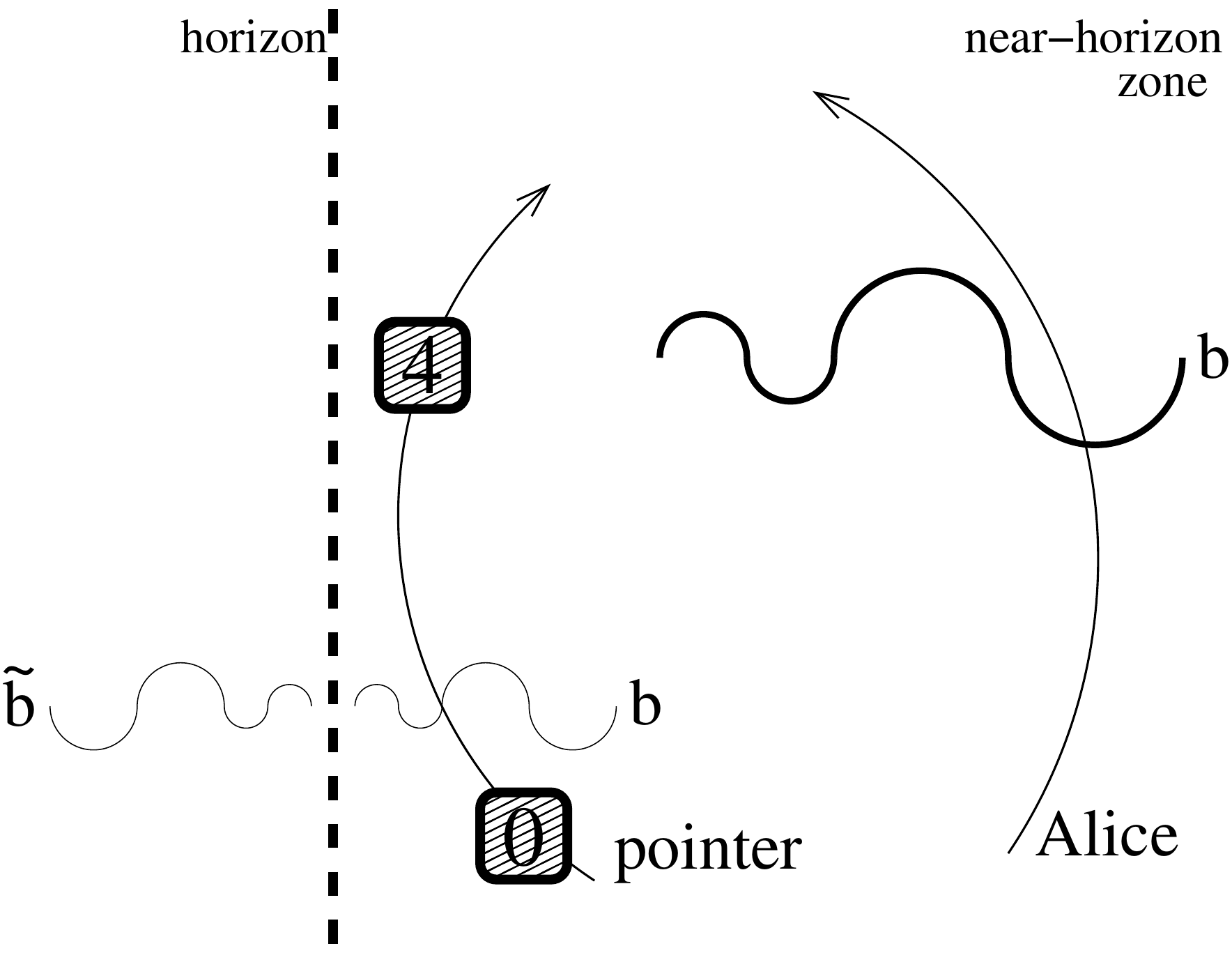}
\caption{A pointer interacts with an outgoing wavepacket in the zone. The equivalence principle then demands that Alice must {\em not} find the vacuum at the horizon. But the same exterior state will result if the pointer instead measures the purification of $b$ in the Hawking radiation (not shown). A self-correcting construction of the interior mode $\tilde b$ cannot distinguish between these two processes, and so will produce the vacuum in both cases. Without corrections for interactions, the map will produce a firewall even if only the Hawking radiation interacts. Either way, the horizon is special.}
\label{fig-freeze}
\end{figure}

\paragraph{Frozen vacuum} Now suppose that the pointer does not measure $e_b$. Instead, when the wavepacket $b$ is 1 light-year from the horizon, the pointer measures $b$ in the occupation number basis (Fig.~\ref{fig-freeze}). This interaction, too, results in the state (\ref{eq-2}). If we like we can correlate $b$ to additional pointers or environments $q, r, s, \ldots$. This would increase the number of factors in each term of Eq.\,(\ref{eq-2}) while preserving its structure. It corresponds to classical broadcasting of the measurement: we can trace over any pointer, and obtain a mixed ensemble. In each pure state that is a member of the ensemble, all remaining subsystems will agree on the outcome of the measurement. These are standard properties of quantum measurements.

Suppose that Bob, who hovers near the horizon, looks at one of the pointers and sees that it shows ``4''. Then if he were to measure $b$ directly, he will also find occupation number 4, with unit probability. Now the pointer(s) and Bob disperse.

Nine years later, a clueless Alice happens to fall through the zone without encountering Bob or the pointers. She does encounter the mode $b$, ten light-years from the horizon, as well as $\tilde b$, inside the horizon. She makes no particular measurement but just enjoys the vacuum. After all, her theory of black holes says that $\tilde b$ must be identified with whatever purifies $b$, whether or not Alice controls the purifying system or has any idea where it is. By Eq.\,(\ref{eq-2}), the purification happens to be a subspace of $ep$. The associated donkey map is Eq.\,(\ref{eq-donkey}), and the result is the infalling vacuum (\ref{eq-3}).

But this contradicts the fact that Alice could have met the pointer, or Bob. Indeed, she could have arranged for the earlier measurement herself. It follows that Alice is chronically unable to find anything {\em but\/} the vacuum, even in cases where her own actions should have destroyed the vacuum.\footnote{It is interesting to ask how Alice interprets this bizarre phenomenon. She can choose not access the environment $p$, and there should be a description of the systems she does control, $b$ and $\tilde b$. The donkey map effectively imposes a particular state for $b\tilde b$ at the horizon. This can be viewed a special case of final state quantum mechanics~\cite{AhaBer64,GelHar91}, where certain histories are post-selected to conform to a particular outcome of a final measurement. The vacuum is not imposed anywhere else, so the horizon is special. (Thus, the difficulty noted here suggests that the black hole final state proposal~\cite{HorMal03} is similarly in conflict with the equivalence principle, since the objections of~\cite{GotPre03} force the final state to be effectively imposed already on the horizon, not at the singularity.)} This means that the vacuum near the horizon behaves differently from the vacuum elsewhere. Alice can detect the location of the horizon by her inability to produce particles. This violates the equivalence principle. 



\paragraph{A more complicated rule?}

Perhaps the trivial donkey map, Eq.~(\ref{eq-olddonkey}), should be applied to a state such as (\ref{eq-2}), which resulted from an interaction with $b$. This map does not involve the pointer, so it could be traced out and the correct (mixed, nonvacuum) state for $b\tilde b$ would obtain. But the state (\ref{eq-2}) would also have resulted if the pointer had measured $e_b$ instead of $b$. This corresponds physically to interactions of the Hawking radiation with some environment. A firewall would not be acceptable in this situation, so we must demand that the donkey map for the state (\ref{eq-2}) is given by Eq.\,(\ref{eq-donkey}). But then the vacuum survives even when it is probed directly.

Perhaps the donkey map (\ref{eq-donkey}) should be used if the state (\ref{eq-2}) resulted from interactions of the Hawking radiation with an environment, but the trivial map (\ref{eq-olddonkey}) should be used if the same state resulted from a measurement of $b$. But this distinction is not well-defined, since in general all systems interact. For example, starting from state (\ref{eq-1}), suppose that Alice transports both $p$ and $e_b$ coherently to the zone. She can arrange for the state (\ref{eq-2}) to emerge from an interaction between all three systems, such that no single system evolves freely during the interaction.  I do not believe that a sensible rule can be formulated that discriminates between $e_b$-$p$ and $b$-$p$ interactions (while allowing for intermediate possibilities), before deciding whether and in what form to apply the donkey map. This is the challenge.

\paragraph{Large systems}

In a more realistic model, the early radiation would consist of a very large number of degrees of freedom, and the purification of $b$ would be a highly scrambled subsystem that may be difficult to access directly~\cite{HarHay13}.  So let us consider coupling the pointer to a random Hawking particle $e$. This is not the same thing as coupling it to $e_b$, and so it will not result in the same state as coupling the pointer directly to $b$. 

One might wish to declare that coupling a pointer to either $b$ or $e_b$ destroys the vacuum---the donkey map should not be adjusted to include the pointer---but after measuring $e$ the map would be adjusted to preserve the vacuum~\cite{MalSus13}. Harlow [private communication] has noted that there exists an interpolation between computationally simple ($e$) and complex ($e_b$) measurements, since the latter can be built out of many simple gates. Here I argue that even if a complexity-based distinction could be upheld, the adjustment (or not) of the donkey map would additionally have to depend on {\em which simple modes} are measured, $e$ or $b$, and to confront intermediate cases.

To keep the equations legible, I will truncate occupation numbers to $0,1$. Individual modes are thus represented by qubits. The vacuum is represented by the state 
$|0\rangle_{b \tilde b}\propto |0\rangle_b |0\rangle_{\tilde b} + |1\rangle_b |1\rangle_{\tilde b}$.
Let $e$ be an easily measured Hawking particle, and let $f$ be the rest of the early Hawking radiation, consisting of an enormous number of qubits. As before, $p$ is a pointer that can be coupled to the outgoing mode $b$, or to $e$. Initially, the state is
\begin{eqnarray}
|\psi\rangle_{befp}\propto |0\rangle_p  \otimes 
\left[ |0\rangle_b \right. & \otimes & \left(|0\rangle_e |\alpha\rangle_f + |1\rangle_e |\beta\rangle_f\right) +  \nonumber \\
|1\rangle_b  & \otimes & \left. \left(|0\rangle_e |\gamma\rangle_f + |1\rangle_e |\delta\rangle_f\right) \right]
~,
\label{eq-1big}
\end{eqnarray} 
where $|\alpha\rangle,\ldots,|\delta\rangle$ are four mutually orthogonal states in the vast system $f$. This state satisfies the condition that in typical states, the small subsystem $b$ must be maximally entangled, and so must the small subsystem $e$. 

First suppose that the pointer interacts with $b$, resulting in the state
\begin{eqnarray} 
|\psi\rangle_{befp} & \propto &  
|0\rangle_b |0\rangle_p\otimes \left(|0\rangle_e |\alpha\rangle_f + |1\rangle_e |\beta\rangle_f\right) \nonumber \\
 & + & |1\rangle_b |1\rangle_p \otimes \left(|0\rangle_e |\gamma\rangle_f + |1\rangle_e |\delta\rangle_f\right) ~.
\label{eq-pb}
\end{eqnarray} 
If we did not trace over the pointer, the donkey map would restore the vacuum, in violation of the equivalence principle. Hence, we must trace over $p$. This results in a mixed state for $bef$, 
\begin{eqnarray} 
|\phi_1\rangle_{bef} & \propto & |0\rangle_b \otimes \left(|0\rangle_e |\alpha\rangle_f + |1\rangle_e |\beta\rangle_f\right) ~~~[50\%]\nonumber \\
|\phi_2\rangle_{bef} & \propto & |1\rangle_b \otimes \left(|0\rangle_e |\gamma\rangle_f + |1\rangle_e |\delta\rangle_f\right) ~~~\,[50\%]~,
\label{eq-mixb}
\end{eqnarray} 
to which no donkey map can be applied. The vacuum is destroyed by the measurement, as required.

Now suppose that the pointer interacts with $e$ instead, resulting in the state
\begin{eqnarray} 
|\psi\rangle_{befp} & \propto &
|0\rangle_b \otimes \left(|0\rangle_e |0\rangle_p |\alpha\rangle_f + |1\rangle_e |1\rangle_p |\beta\rangle_f\right) \nonumber \\ & + &
|1\rangle_b  \otimes \left(|0\rangle_e |0\rangle_p |\gamma\rangle_f + |1\rangle_e |1\rangle_p |\delta\rangle_f\right) ~.
\label{eq-pe}
\end{eqnarray} 
Qualitatively, the choice is exactly the same as for Eq.\,(\ref{eq-pb}): we can keep $p$ or trace over it. If we traced over $p$, we would obtain a mixed state for $bef$, with equal probability for the two states
\begin{eqnarray} 
|\phi_1\rangle_{bef} & \propto & |0\rangle_e \otimes \left(|0\rangle_b |\alpha \rangle_f+ |1\rangle_b |\gamma \rangle_f\right) ~~~[50\%]\nonumber \\
|\phi_2\rangle_{bef} & \propto & |1\rangle_e \otimes \left(|0\rangle_b |\beta \rangle_f+ |1\rangle_b |\delta \rangle_f\right) ~~~\,[50\%]~.
\label{eq-mixe}
\end{eqnarray} 
No single map from $bef$ to $\tilde b$ can convert this mixed state into a pure state such as the infalling vacuum. So in this case, the desired outcome forces us to retain $p$ and apply the donkey map to the full system. 

But there appears to be no well-defined distinction between interactions of $b$ and $e$ with the environment. As discussed earlier, $e$ and $p$ can be coherently transported to $b$ and all three bits made to interact simultaneously. Because $e$ is a simple Hawking particle, no computation is required for its extraction from $ef$, so the concerns of~\cite{HarHay13} do not apply. Moreover, $e$ is only one quantum, so its backreaction in the zone is negligible.

\paragraph{Quantum error correction} There is one difference between the mixed states (\ref{eq-mixb}) and (\ref{eq-mixe}): In the ensemble (\ref{eq-mixb}), $b$ is pure in each of the two outcomes, so no donkey map exists for any of the outcomes. But in the ensemble (\ref{eq-mixe}), $b$ is maximally entangled in each outcome. Hence, a donkey map could be defined for each outcome, from $ef$ (or just from $f$) to $\tilde b$, so as to obtain the vacuum $|0\rangle_{b\tilde b}$. The map now depends on the outcome of the measurement, in addition to its inevitable dependence on the overall out-state, Eq.\,(\ref{eq-pe}). This dependence can be removed by applying a unitary operation $U_f$ to $f$ that rotates the purification of $b$ to a special bit $f_1$. One can then view the donkey map as a single map acting only on $f_1$, not on the remaining factor whose role is merely to purify $e$ (and to discriminate the outcomes after $e$ is measured).

This suggests that the redundancy of the purification of $b$ can be exploited to reconstruct $\tilde b$ even after interactions have taken place, in a spirit reminiscent of quantum error correction~\cite{VerVer12,MalSus13}.\footnote{Independently of the obstruction identified here, this analogy seems inappropriate. Error correction is useful in quantum computing when we do not know (a) whether an error occurred and (b) what error occurred, but we can (c) be reasonably confident that the number of errors is limited. Moreover, (d) an ancillary system must be available to absorb the error. None of the above criteria naturally carries over to the black hole case, where we can grant complete knowledge of the unitary evolution and full control over the radiation and any environment. Consistency cannot require additional systems such as an ancilla, and it is not clear what existing subsystem could play this role. If interactions, or even the unitary out-state itself, are treated as errors, then these errors need not be limited to a small code subspace, since all of the radiation can interact.---Moreover, error correction, by construction, will not extract any information from the computational bits. Thus, if a physical error correction protocol automatically undid the results of any interaction with Hawking radiation, then the out-state could not be measured by any apparatus, so information would be operationally lost.}
However, this breaks down in the generic case where all of the radiation interacts with a much larger environment. An untouched portion of the original system, such as $f$ above, contains an approximate purification of $b$ only if $f$ constitutes more than half of the qubits of the total system. For an old black hole, the radiation alone constitutes more than half of the system, so a decoherent measurement of the radiation destroys {\em all\/} possible purifications of $b$.  Then $b$ will be pure in every member of the resulting ensemble.  The only option for recovering the vacuum is to apply a donkey map to (at least half of) the entire system consisting of black hole, the radiation, and the vastly larger environment it has interacted with.

But then we are back to the problem discussed above: why not apply the same map to recover the vacuum if $b$ has also been measured? Again, there is no sharp distinction between interactions that only involve most of the radiation, and interactions that in addition involve $b$. For example, before the entire radiation runs into galactic dust, a few quanta from the Hawking radiation can be transported to the zone where they interact both with $b$ and with some environment, simultaneously. By dialing the the relative strength of the $b$-$p$ and $e$-$p$ interactions in the zone, we can interpolate between interactions that must be corrected, and interactions that must not be corrected.


To summarize, it is unclear how to formulate any criterion that would allow the reconstruction of the infalling vacuum to depend on whether the zone was involved in interactions. Either we {\em always\/} retain the environment that the zone and radiation have interacted with, and we apply a donkey map to this large system to construct $\tilde b$. Then we obtain the vacuum when we expected a firewall. Or we {\em always\/} ignore the environment; then we get a firewall when we expected the vacuum. The latter choice is standard; but with either choice, the horizon is special.

\paragraph{Acknowledgments} I am indebted to J.~Maldacena and L.~Susskind for extensive discussions. This work was supported by the Berkeley Center for Theoretical Physics, by the National Science Foundation (award numbers 1002399, 0855653 and 0756174), by fqxi grant RFP3-1004, by ``New Frontiers in Astronomy and Cosmology'', and by the U.S.\ Department of Energy under Contract DE-AC02-05CH11231.

\bibliography{all}
\end{document}